# The Ancient Astronomy of Easter Island: The Mamari Tablet Tells (Part 1)

Sergei Rjabchikov[1]

[1]The Sergei Rjabchikov Foundation - Research Centre for Studies of Ancient Civilisations and Cultures, Krasnodar, Russia, e-mail: srjabchikov@hotmail.com

## Abstract

The ancient priest-astronomers constantly watched many heavenly bodies. The record about Halley's Comet of 1682 A.D. has been decoded completely. Good agreement between it and the results of European astronomers is seen. The records about Halley's Comet of 1835 A.D. as well as about the sun, the moon, Mars and Saturn have been deciphered as well. The obtained information is the basis in order to understand some aspects of the bird-man cult.

## Introduction

The civilisation of Easter Island is famous due to their numerous ceremonial platforms oriented on the sun (Mulloy 1961, 1973, 1975; Liller 1991). One can therefore presume that some folklore sources as well as *rongorongo* inscriptions retained documents of ancient priest-astronomers.

## The Faint Echo from Rapa Nui

Routledge (1998: 249) says about the *rongorongo* board called *kohau-o-te-ranga* (the Mamari tablet):

It was the only one of the kind in existence, and was reported to have been brought by the first immigrants; it had the notable property of securing victory to its holders, in such a manner that they were able to get hold of the enemy for the "ranga" – that is, as captives or slaves for manual labour.

Consider the record on the Mamari tablet (C), see figure 1.

Ca 13: 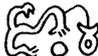

Figure 1.

Ca 13: **62 39 28** *toa ranga* the warriors (of the Tupa-Hotu tribe) are captives

## The Report about Halley's Comet of 1682 A.D.

Consider the record on the same tablet, see figure 2.

Cb 9: (a) 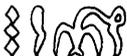

(b) 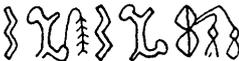

(c) 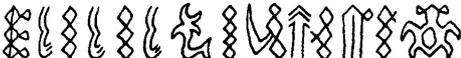

Figure 2.

Cb 9: (a) **17-50 44** *Tei ta(h)a*. The frigate birds (sooty terns figuratively) were visible (= appeared).
(b) **52 143 24 52 143 17 17-17** (or **120-120**, or **67-67**) *Hiti Paupau ari, hiti Paupau tea, teatea* (or *ngingi*, or *pipi*). The bright Halley's Comet (1P/Halley) appeared, the clear Halley's Comet appeared, (it) shone (or was bright, etc.).
(c) **17 43 17 43 17 43 11 17 61 17 4/33 17 26 17 4 …** *Te ma, te ma, te ma Maho (= Mango), te hina, te atua/ua, te Ma, te ha…* (The following data are here:) the motion, the motion, the motion in the Virgo constellation, (the calendar counting:) the moon, the 13th moon, the Bright Light (the deity *Ma = Maa*, the symbolism of the god *Makemake* < *Ma-ke*), the fourth (month)…

Glyph **143** is the sign of Halley's Comet (with the two glyphs **35** *Pa*, *Pau* 'the star *Pau*'). Glyph **43** *ma* depicting the leg reads *ma* (it also reads *vae* 'leg; foot' seldom), cf. Maori *ma* 'to come; to go.'

Let us investigate a Rapanui chant (Campbell 1999: 217; the translation in Rjabchikov 2013a: 6):

| | |
|---|---|
| *Ka moe nga pua.* | Eggs (*pua*, *hua*) slept. |
| *Mo roto i te tama ere* | (They) were inside (nests) for a young man (*tama ere*) |
| *mo hiki, mo turu* | for the elevating (*hiki*), for the coming down |
| *ki te honga'a pua.* | towards the nests of the eggs. |
| *Teitei Renga o nga manu;* | (It was the god) *Teitei Renga* (the sun god *Makemake*) of many birds; |
| *Keu Renga.* | (it was the god) *Keu Renga* (the sun god *Makemake*). |

In this text a servant *hopu* who looked for an egg of sooty terns *manu-tara* on the islet Mitu Nui in the month *Hora-nui* (September chiefly) is described. According to Métraux (1940: 333ff; 1957: 130), during that annual festival such a young man fetched the first found egg from the islet to the ceremonial village of Orongo. Then his master, a victorious warrior, received this sacramental egg (the incarnation of the god *Makemake*) and was proclaimed as the new bird-man.

The name *Teitei Renga* means 'The arrival or elevation of the Yellow Colour (the sun).' This colour was the notation of the sun in several Rapanui place names (Rjabchikov 1998). The name *Keu-Renga* signifies 'Shelter of the Yellow Colour (= The Nest of Sacred Egg).'

In the Rapanui folklore text known as the Creation Chant (Métraux 1940: 320-322) the character *Tei* (Highness) is rendered. Rapanui *teitei* 'to grow; to increase; to raise; to elevate; height' is comparable with Maori *teitei* 'summit, top,' Tahitian *faateitei* 'to raise,' Mangarevan *akateitei* 'to raise up,' Tuamotuan *fakateitei* 'to raise,' and Samoan *te'i* 'to be rising (of the tide)' < PPN *\*teki* 'to rise; to arrive.'

The god *Hiku-nene-ko-mo-toi-pua* (cf. Rapanui *hiku* 'tail') was presented in the legend about a local war (Ibid., pp. 382-383). In my opinion, it was the designation of Halley's Comet of 1682 A.D. Here and everywhere else, I use the computer program RedShift Multimedia Astronomy (Maris Multimedia, San Rafael, USA) to look at the sky above Easter Island.

The term *nene* corresponds to Rapanui *nenenene* 'nice,' hence the expression *hiku nene* means 'nice tail (comet).' The term *Mo-toi* is *Motohi* or *Omotohi* (*Ma-tohi*, *Mo-tohi*), the name of the 18th moon of the lunar calendar. It is comparable with the name of the moon goddess *Toitoi*. The new moon was on September 1, 1682 A.D., hence *Motoi* = September 18 or 17 (if the duration of the lunar month was shortened), 1682 A.D. It is common knowledge that on September 17 the comet was bright, its tail was around $16^0$ long (Belyaev and Churyumov 1985: 37). Here term *pua* means 'full' (= the lunar phase *Motohi*). In conformity with Kronk (1999: 373), the comet was in Virgo on September 7 and 10.

I have chosen September 7, 10, 17 and 18 as the dates for the computer simulations. The comet was in Virgo in those nights.

In the Rapanui rock art this heavenly body is depicted on a *moai* pedestal stone at Hanga Ohio (Lee 1992: 94, figure 4.84; the interpretation in Rjabchikov 1997: 204). The comet resembled a club (with glyph **35** *Pau*). The turtle/bird sign of the Pleiades (M45; NGC 1432) was the celestial marker at the end of August (see Rjabchikov 2016a: 13).

During that terrible war between the western tribal union Hanau Momoko (Moko, Tuu, Miru etc.) and the eastern tribal union Hanau Eepe (Hotu[-Iti], Tupa[-Hotu] etc.) the first won. By the way, the terms *Eepe* and *Hotu* were synonyms ('Fats' and 'Bearing fruits' = 'The abundance and wealth'). The decisive battle occurred in September 1682 A.D. Since that event, the bird-man cult of the Tupa-hotu tribe (the descendants of the builders of *tupa* towers) was adopted by the western tribes.

# The Report about Halley's Comet of 1835 A.D.

Consider the record on the Great Washington tablet (S), see figure 3.

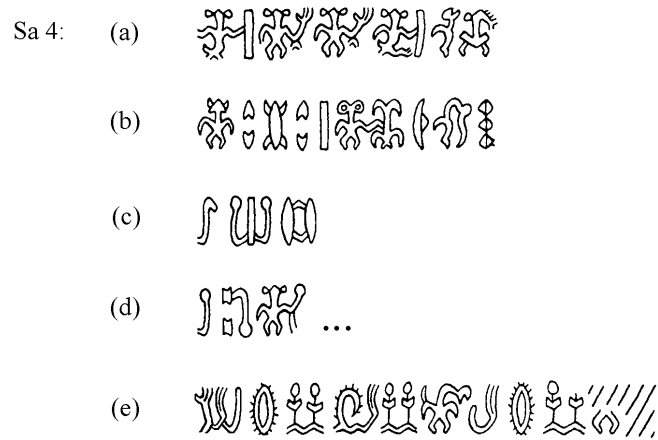

Sa 4: (a)

(b)

(c)

(d)

(e)

Figure 3. (corrected)

Sa 4: (a) **6-4 6-15-6-15 6-4 19 49** *Hotu Horahora, hotu ki (ariki) mau*. (The month) *Horahora = Hora-nui* (September/October) produced fruits (including eggs), the fruits (including eggs) were for the king.
(b) **6-21 29-28-29 4 60 80** (or **23 44**) **3 44b-17** *Hakarungaru atu Mata ui, Hina, Tua tea*. Mars (The visible Eye literally), the Moon, Venus (Saturn indeed) were elevated (in the sky).
(c) **35 4-4-4 30-149-149-30** *Pau atua, atua, atua ana hatuhatu ana*. A great deity (god-god-god literally), (the star) *Pau* (Halley's Comet of 1835 A.D.; e.g., on October 19), shone very brightly.
(d) **4 29-4 6-4** (a damaged segment with several unclear glyphs) *Atua Ruti hotu...* The god of the month *Ruti* (November chiefly) yielded harvest…
(e) (a damaged segment) **15 58 4 14 151 14 58 151 62 58 14 151 62** (a lost segment) … [SUGAR CANES] *roa, tahi tia; hau* SUGAR CANES*; hau tahi* SUGAR CANES *toa tahi; hau* SUGAR CANES *toa...* ... [Sugar canes] grew, the first (fruits) were cut down; the royal possession (ownership, receipt) of the sugar canes; the first royal possession of the first sugar canes; the royal possession of the sugar canes…

    The initial drawing of the glyphs was corrected on the basis of the set of photos that is available online.[2] The text has partially been investigated (Rjabchikov 1999: 18).
    The Rapanui calendar terms *Hora-iti* (August chiefly) and *Hora-nui* (September chiefly) are well known (Métraux 1940: 51). The form *Horahora* (September or August-September chiefly) is registered in the folklore text "*Apai*" (Rjabchikov 1993: 132). In the Maori beliefs, the planets Venus and Saturn were the husband and wife (Best 1922: 35).
    Old Rapanui *roa* means 'long; large; to grow,' cf. Rapanui *roa* 'long; large,' *roroa* 'to grow tall.' Old Rananui *ti* and *tia* mean 'to pierce; to cut; to dig; to beat; stick,' cf. Rapanui *ivi tia* 'sewing needle,' *tiaki* 'to dig a hole in the ground' (< \**tia ki*), Mangarevan *tia* 'to pierce, to stick in' and *tiatia* 'to prick; to stick in a thing with a point.'
    The form *rungaru* (glyphs **29-28-29**) is incomplete reduplication of the word *runga* (top), cf. the word *nihoni* (indented?) put down in Manuscript E (Barthel 1978: 324, 326) instead of the word *nihoniho*. Cf. also Rapanui *taheta* 'fountain; spring' and *tahe* 'to flow.' Glyphs **6-21** denote the causative prefix *haka-*. Glyphs **14** *hau* (royal possession) correlate with Old Rapanui *hau* 'king' and Rapanui *hauhau* 'ownership.'
    In conformity with Belyaev and Churyumov (1985: 76), Halley's Comet was slightly brighter than the star Aldebaran (α Tauri) on October 19, 1835 A.D. That event could be described in segment (b). It happened at the end of the month *Tangaroa-uri*. The new moon of the next month *Ruti* occurred on October 22, 1835 A.D. The month of harvesting sugar canes was October (Barthel 1978: 52).

---

[2] See those photos: <http://collections.nmnh.si.edu/anth/pages/nmnh/anth/Display.php?irn=8010185>.

# The Cross-readings in Different *Rongorongo* Inscriptions Including Astronomical Ones

1. Let us consider the parallel records on the Great Santiago (H) and Great St. Petersburg (P) tablets, see figure 4.

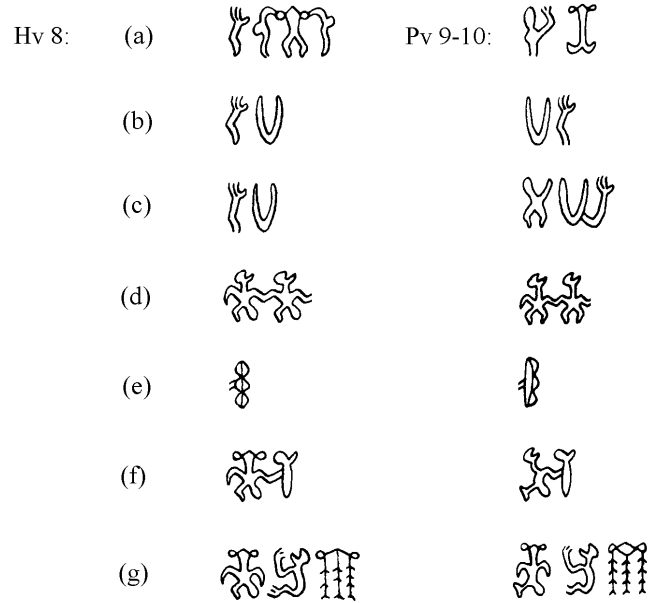

Figure 4.

Hv 8: (a) **15 75 44** (b) **15 61** (c) **15 61** (d) **6-6** (e) **17** (f) **6 27** (g) **6 62var** [= sic! = it was copied from another tablet with glyph **62** in the text] **24-24-24** (a) *Roa ko Taha,* (b) *roa Hina,* (c) *roa Hina.* (d) *Haha,* (e) *tea* (f) *a Rau (Rahu)* (g) *a Tonga Ari(k)i.* (a) The Frigate Bird (the faint solar rays of early dawn; the sun in the morning) rose, (b) the moon goddess *Hina* (the moon) appeared, (c) the moon goddess *Hina* (the moon) appeared. (f) *Rau (Rahu)* (d) (e) watched (them) (g) at Tongariki (Tonga Ariki).

Pv 9-10: (a) **73 15 56-50** (b) **61 15** (c) **84 61 15** (d) **6-6** (e) **18** (f) **6 27** (g) **6 62var** [= sic! = it was copied from another tablet with glyph **62** in the text] **24-24-24** (a) *He roa poi*, (b) *Hina roa*, (c) *ivi Hina roa*. (d) *Haha,* (e) *tea* (f) *a Rau (Rahu)* (g) *a Tonga Ari(k)i.* (a) The morning began (rose literally), b) the moon goddess *Hina* (the moon) appeared, (c) the ancestress, the moon goddess *Hina* (the moon), appeared. (f) *Rau (Rahu)* (d) (e) watched (them) (g) at Tongariki (Tonga Ariki).

    Astronomically, the new moon rose (and rises) during the sunrise. The waxing (full, waning) moon rose (and rises) later and later each next day.
    In the decoded records the priest-astronomer and scribe *Rau* or *Rahu* (*Arahi* < *Rahi* < *Rahu*, *Arohio* < *Rohi* < *Rohu*; *Rau*, *Rou*) from the ceremonial platform Tongariki (Tonga Ariki) who watched the rising of the sun and the moon is mentioned. The prototype of the inscriptions was borrowed from lost local astronomical diaries; it was used in the royal *rongorongo* school. The ancient scholar *Arahi* (*Arohio*) was a friend of king *Nga Ara* (Routledge 1988: 249).
    Glyphs **15** *roa* mean 'to grow; to rise' in both records. In the Old Rapanui language (it is apparent from folklore texts) the prime word order was such: VSO or SVO (cf. Chapin 1978; Fedorova 1978a; Alexander 1981). The name of *Rau* (*Rahu*, *Rou*, *Rohu*) is introduced by the particle *a* (glyph **6**), see other examples of the usage in figure 7.
    In the first record the wordplay is possible: cf. Old Rapanui *taha* 'frigate bird' and Rapanui *taha rangi* 'horizon' (cf. also the term *taha* 'ditto' in Henry 1928: 165). In accordance with a Maori myth (Best 1925: 312), the goddess *Hine* was designated as *tipuna* 'ancestress,' cf. Maori *tupuna*, *tipuna* 'ancestor.'
    In segment (a) of the second record the verb *roa* (to grow; to rise) is introduced by the particle *he* (glyph **73**), see other examples of the usage in figures 8 and 9.

2. Consider the record on the Aruku-Kurenga tablet (B), see figure 5.

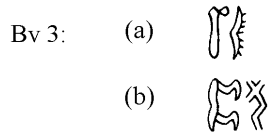

Figure 5.

Bv 3: (a) **56-50** (b) **50-29-70** (a) *poi*; (b) *hi rua, pua* (a) the morning; (b) the sun: the set (and) the rising

The combination of glyphs **50-29-70** is repeated many times on this board (Rjabchikov 2013b: 8, figure 7). It means that the ancient astronomers of Easter Island watched the rising and setting sun constantly. In this context the usage of the term *poi* (morning) was quite natural.

3. Consider the record on the Great Santiago tablet, see figure 6.

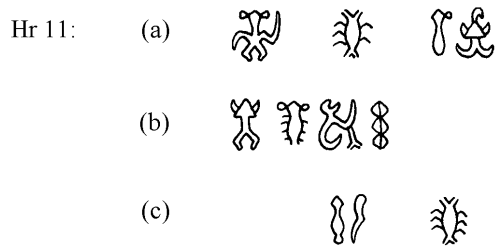

Figure 6.

Hr 11: (a) **6 77 56-50/ 35/21** (a) *Ha pure Poi puoko*. *Poi* [The Morning] gave the *mana* through (his) skull.
(b) **21 56-6 15 17** *Ko Po-a roa, tea*. (It was) the great *Po-a* [The Dawn] (= *Poi*) of the eastern tribes (= the ruler of the high origin as the ruler of the western and eastern parts of the island in this context).
(c) **56-50 77** *Poi pure*. *Poi* [The Morning] gives the *mana*.

The name *Poi* came from Old Rapanui *poi* 'morning,' cf. Rapanui *poipoi* 'ditto' < \**po hi* 'night, solar rays.' This hero was known as *Poia*, *Poie* and *Poio* (cf. the traditions in Thomson 1891: 529-531; Métraux 1940: 74-84; Felbermayer 1971: 51-63). He was a chief of the Miru tribe and participated in the war (wars) between that tribe (the tribal union Tuu) and the tribe Tupa-Hotu (the tribal union Hotu-Iti). *Poia* once built a large double canoe which he called *Tua Poi*. The last name means '(The chief) *Poi* (= *Poia*) from the western part (the back literally) of the island' (see also Blixen 1973: 7-8). One of his brothers who survived in the war was called *Taka Aure* or *Taku-hau-uri*. Perhaps it was *Taka (h)au uri* '*Taka* (wearing) the black hat (the royal hat made of the black feathers of frigate birds),' cf. also Old Rapanui *hau* 'king.' Hence, one can presume that it was *Mataka roa* (The Great *Mataka* = *Taka*), a western leader during the war between Miru (Moko, Hanau Momoko) and Tupa (Tupa-Hotu, Hanau Eepe).

The full name of *Poie* (*Poia*) was *Poie-nuinui-a-tuki* (the great *Poie*, son of *Tuki* or the great *Poie*-the killer). One Rapanui song was dedicated to *Poie* (Barthel 1962: 848). My own decipherment of that text has shown that he was the first warrior (*i mua ia koe taau taua*) armed with a long regal spear (*vero*, *vero*, *vero*, *vero*: the term *vero* 'spear' was repeated four times). He had the high title *hatu paki hatu paki* (the great progenitor-seal; it was the symbolism of *Tangaroa*, the prime deity of the Miru tribe). He had the supernatural power of the sun (*mana nia* = *mana ngia*) that had been received at the religious centre (house-temple) at *Tuu Tapu*. He possessed small and big *tohu* (shelters, houses, caves), small and big *mara* (cultivated fields), cf. Mangarevan *tohu* 'to hide one's self' and *mara* 'cultivated field.' He received the big fish (*ika nui*), the big fish *pei* (*pei nui*) [*pe hi* < *poo hi* 'the angled *poopoo* fish'?] and the tuna fish (*kahi*). So, he was the great ruler (king) in fact.

As can be seen in figure 6, the skull of this hero provided the growth of the plants (the increasing of the number of eggs, etc.). The name of *Po-a* (*Poie*) is introduced by the particle *ko* (glyph **21**).

4. Consider the record on the Small Santiago (G) tablet, see figure 7.

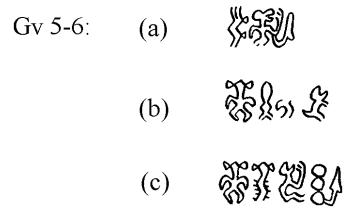

Gv 5-6:     (a)

            (b)

            (c)

Figure 7.

Gv 5-6: (a) **52 19 102** *Hiti ki ure:* (King) *Hiti* informed (*ki*) (his) genealogy (*ure*):
(b) **6 73 62-19** *A e Toki*, (King) *Toki*,
(c) **6 56-6 17 102** … *a Po-a tea ure* a son (*ure*) of (king) *Po-a* [The Dawn] (= *Poi*) of the eastern tribes (= of the royal origin in this context)…

Hence, king *Hiti* (The Appearance, Lifting, Rising) was son of *Toki* (The Adze, Axe), who was son of *Po-a* = *Poie* (The Day-break, Morning). In the Rapanui genealogies of Rapanui supreme rulers (per Métraux and Jaussen) the first two kings are called *Ko te Hiti rua nea* (*Hiti a ua a nea*) and *Ko te pu i te Toki* (*Tu* [= *Tuu*] *pu i te Toki*) (Métraux 1940: between 90-91, table 2). The name and title of the first mean '*Hiti*, the second or next (ruler) here,' cf. Rapanui *na*, *nei* 'here,' and the name and title of his father mean '*Toki*, ruler (*pu*) of the tribal union Tuu,' cf. Maori *pu* 'ruler, king.' The real chronology of the ancient Rapanui history is reflected in this brief list of the kings (Miru, the new elite class).

The name of *Po-a* is introduced by the particle *a* (glyph **6**). The particles *a* (glyph **6**) and *e* (glyph **73**) introduce the name of *Toki*.

5. Consider the record on the same tablet, see figure 8.

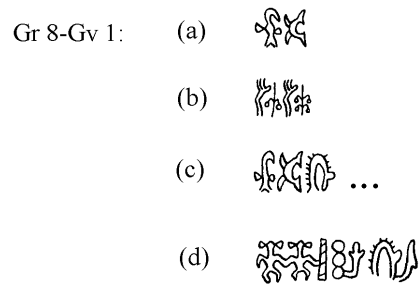

Gr 8-Gv 1:  (a)

            (b)

            (c)        …

            (d)

Figure 8.

Gr 8: (a) **73 35 8** *He pau Matua*. (King) *Hotu-Matua* was buried.
(b) **15 92** (cf. **25var**) **15 92** (cf. **25var**) … *Roa ere (hua), roa ere (hua)*… (The magico-religious consequences were such:) flowers/fruits grew, flowers/fruits grew…
(c) **73 35 8 14 102** … *He pau Matua hau ure*… King-fertiliser *Hotu-Matua* was buried…
(d) **6-6-4 17 25 14 102-102** *Hohotu te hua ureure*. (Plants) bore fruits due to the king-fertiliser.

A fragment of the vast text about the burial of king *Hotu-Matua* has been decoded (cf. Rjabchikov 2009: 6-7). As can be seen in figure 8, the skull of this hero provided the growth of the plants (the increasing of the number of eggs, etc.). According to Manuscript E (Barthel 1978: 237), the skull of king *Hotu-Matua* had the great *mana*. The men of different tribes wanted to steal the buried corpse of the king per the legend "The obsidians of Hare o Ava (Burial of Hotu Matu'a)" (Englert 2002: 70-71).

Glyphs **73 35** *he pau* stand at the beginning of segments (a) and (c). *He* is the verbal particle in both cases.

6. Consider the parallel records on the Great St. Petersburg, Great Santiago and Small St. Petersburg (Q) tablet, see figure 9.

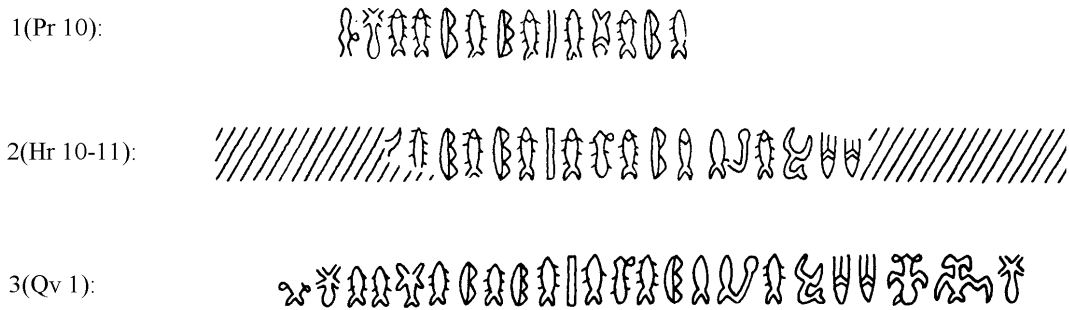

Figure 9.

1 (Pr 10): **73 45 12 12 18 12 18 12 4 12 16 12 18 12** … *He pu ika, ika tea, ika tea, ika atu, ika kahi, ika tea, ika…*
2 (Hr 10-11) (a damaged segment) **11 12 18 12 18 12 4 12 56 12 18 12 12 4 12 2 1-1** (a damaged segment) … *mango (niuhi, pakia* etc.*), ika tea, ika tea, ika atu, ika poo (= poopoo), ika tea, ika, ika atu, ika Hina Tikitiki…*
3 (Qv 1): **73 45 12 12 11 12 18 12 18 12 4 12 56 12 18 12 12 4 12 2 1-1 68 44-45** *He pu ika, ika, mango (niuhi, pakia* etc.*), ika tea, ika tea, ika atu, ika poo (= poopoo), ika tea, ika atua, ika Hina Tikitiki, honu tapua.*

The texts tell of increasing the number of all fishes, dolphins (sharks etc.) and sacramental turtles (at least in the last case). Glyphs **73 45** *he pu* stand at the beginning of fragments 1 and 3. *He* is the verbal particle in both cases. Without going into details the three sentences mean '(Sea creatures) were numerous in the summer (*Hina Tikitiki* = the moon associated with the hot sun).'

According to Popova (2012), Old Rapanui *pu* 'to produce' (Métraux 1940: 320-322) is cognate with Maori *pupu* 'to spring up; to grow; to appear' < **pu-pu*, *pukahu* 'abundant' < **pu kau* (*kau* 'ancestor'), and Hawaiian *pu* 'to come forth from.' Cf. also Rapanui *pupu* 'to collect; to accumulate.'

The main glyph is sign **12** *ika* 'fish' repeated many times. Glyph **18** *tea* (*atea*) denotes the king's fish *kotea* (Métraux 1937: 50). I have compared that name with the Marquesan fish names *ko'otea* = *'otea* (Randall and Cea 1984: 14). Glyph **16** *kahi* denotes the tuna fish, glyph **4** *atu* the *atu* fish, and glyph **56** *poo* the *poopoo* fish. Glyph **11** *mango (niuhi, pakia)* denotes the shark, dolphin, seal, whale etc. Glyphs **68 44-45** *honu tapua* denote the sacred turtle.

In fragment 1 the names of the tuna fish and the *atu* fish are placed alongside each another as they are placed (but in the reverse order) in a chant about the power of the king (Métraux 1937: 52-54).

The names of the tuna fish and the *poopoo* fish substitute for one another in the records, cf. fragment 1 and the rest. In compliance with the legend "Prohibition of tuna and po'opo'o fishing" (Englert 2002: 200-201), king *Hotu-Matua* forbade catching the tuna fish and *poopoo* fish during six months of the winter; they were allowed after the month *Hora-nui* (September chiefly) only.

7. Consider the record on the Aruku-Kurenga tablet, see figure 10.

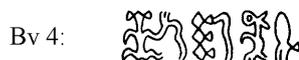

Figure 10.

Bv 4: **69-50 17-50 19 43/73** *Makoi tei ki maea*. The *makoi* trees (Thespesia populnea) grew near stones.

Métraux (1940: 17) says about this plant:

The only *makoi* trees that I saw were growing on the side of the high cliffs of Poike where they cannot be reached, though I was able to collect many dry fruits at the base of the cliffs. These were subsequently identified in Paris as the fruit of Thespesia populnea.

The word *maea* 'stone' is written with the help of glyphs **43-73** *ma-(h)e*. This word came from the expression **ma ea* 'for the lifting.'

8. Consider the record on the Tahua (A) tablet, see figure 11.

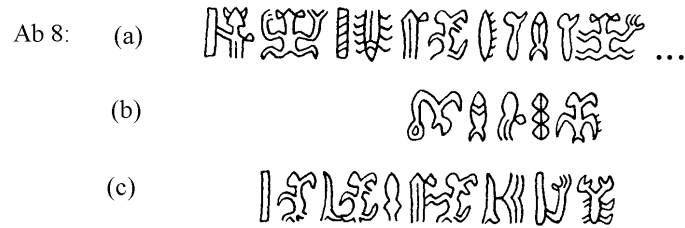

Figure 11.

Ab 8: (a) **4-26 21 69 4 1 26-19 46 27 12 27 69 …** *Timo ako: Moko atua Tiki-Maki, naa, ro(h)u ika, ro(h)u moko…* The pupil is carving (writing): The god *Tiki-Makemake* vanished, (he) hid himself, fishes (appeared) in the darkness (in the water)…
(b) **65/31 12 43/73 17 44** *Rangi Maki ika maea. Te taha.* (The god) *Makemake* called, (another) fish (appeared) from the stone. He was alone (cf. Rapanui *tahanga* 'alone; sole; only,' *noho tahanga* 'unmarried').
(c) **4-19-5-19 73 26-19 5-33 4-15 20** *Tukituki e Maki-atua, atua roa Ungu.* The god *Makemake* copulated, the great (moon) goddess 'The Crab' (was born).

According to Barthel (1957: 70), the epithet *Makiki* of the god-creator *Tiki* in the Tuamotuan mythology is comparable with name of the Rapanui god *Makemake*. In the inscription this personage is called *Tiki-Maki* (glyphs **1 26-19** *Tiki Ma-ki*) in segment (a), *Make* or *Maki* (glyph **31**) in segment (b), and *Maki* (glyphs **26-19** *Ma-ki*) in segment (c). This record is preserved in several folklore versions (cf. Métraux 1940: 320-322 (several sentences about *Tiki* in the Creation Chant); Heyerdahl and Ferdon 1965: figure 147; Barthel 1957: 63). I think that this deity was a complicated figure (the god of the solar cult, the god-progenitor, and the god who granted the wealth and abundance in different conditions).

The name of *Maki* (*Tiki-Makemake*) in segment (c) is introduced by the particle *e* (glyph **73**). Again, the word *maea* 'stone' is written with the help of glyphs **43-73** *ma-(h)e*.

9. Consider the record on the Santiago staff (I), see figure 12.

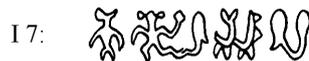

Figure 12.

I 7: **44 6-4 (102 123) 51 48 25** *Taha Hatu a Keu hua.* The frigate bird (the sooty tern figuratively) of the god *Tiki-te-Hatu* as 'The shelters (nests) of the eggs.'

10. Consider the record on the Small Santiago tablet, see figure 13.

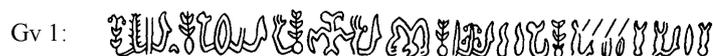

Figure 13. (corrected)

Gv 1: **101 92var (25var) 25-4 (102) 3b** (= the inverted moon glyph **3b** *hina*) **101 27-28-4 (102) 15 101 44 58 (102) 31 101 30 25 (102) 50-50 27 101** (a damaged segment) **9 9 (102) 30 9 …** *O(h)o ARE Huti Ino (Ina), o(h)o, ro(h)u, ngatua, roa; o(h)o, taha tahi Make; o(h)o ana hua Hihi; ro(h)u, o(h)o … niu, niu, ana niu…* Bananas (glyphs **25-4** *huti*) of the variety *Ino* (*Ina* 'Resembling a Crescent') appeared, (they) appeared, were planted (*ngatua*), grew (*roa*) in the great number; the first (fruits) were cut down for (the god) *Makemake*; bananas of the variety *Hihi* (glyphs **50-50**; 'The High Trees' or 'Growing on a Slope') appeared in the great number; nuts appeared in the great number.

Interestingly Old Rapanui (1774 A.D.) *huti* (*futi*) means 'banana' (Langdon and Tryon 1983: 31-32). The banana varieties *maika-puru-ino* and *maika-hihi* were known (Métraux 1940: 156). Old Rapanui *taha* 'to cut' correlates with Rapanui *taha* 'to tear,' Niuean *tafa* 'to cut with a sharp instrument' and Mangarevan *tahataha* 'to cut into pieces.'

11. Consider the record on the Great Washington tablet, see figure 14.

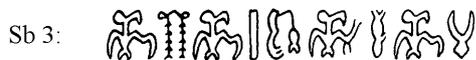

Figure 14.

Sb 3: **44 24-24 44 4 43-12 44 4 9 44 54var** *Taha ariari, taha, tia maika, taha, tia niu, taha kai.* Nuts (cf. Rapanui *tiari* 'nut' < **ti* or *tia ari* 'to cut a nut down') were cut down (*taha*), bananas (glyphs **43-12** *maika*) were cut down (*taha, tia*), nuts (*niu*) were cut down (*taha, tia*), (thus, the different sorts of) the food were cut down.

In the legend "The living remembrance" (Englert 2002: 108-109) such words were said: *he tari te kai: te ura, te ika, te koreha, te toa, ananake te kai* '(the children) brought (them) the food – eels, fishes (and) sugar canes – all kinds of the food.'

12. Consider the record on the Santiago staff, see figure 15.

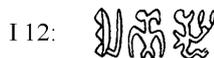

Figure 15

I 12: **18-4 44 72 64** *te atua Taha – Manu mea* the deity 'The Frigate Bird – The Red Bird'

This text has been decoded earlier (Rjabchikov 1993: 135-136, appendix 2, figure 5, fragment 24). It is the description of the sacred bird of the god *Tane* (*Tiki-Makemake*) known as *Manu Mea* (The Red Bird) in the Maori mythology (see Rjabchikov 2016b: 3). Cf. the name of the Rapanui deity *Manu mea* (Métraux 1940: 318). A report of McCoy (1978: 204) about his studies of archaeological sites on the off-shore islet Motu Nui is noteworthy. A heap of earthen red pigment was found behind the stone wall of a cave (site 1-535) associated with the bird-man cult. A frigate bird petroglyph in that cave was outlined with that pigment (Ibid., p. 208, table). That rock drawing is copied (Lee 1992: 74, figure 4.50). I have decoded it. The main signs are such: a frigate bird, a sooty tern (*manu-tara*), and another frigate bird. The single sign consisting of nine lines (cf. Rapanui *iva* 'nine') denotes the legendary homeland Hiva. Both frigate birds (cf. Rapanui *tahataha* 'side; edge') in that picture could reveal the annual transition of the sun during the fourth month *Hora-nui* (September chiefly; the month of the vernal equinox; cf. the small glyph **6** *ha* 'four' near the second frigate bird) from winter to summer, otherwise from death to life.

**Astronomy and Mathematics: the Mystery of the Number Four**

Consider the record on the Great Santiago tablet, see figure 16 (see details in Rjabchikov 2012a: 568-569, figure 8).

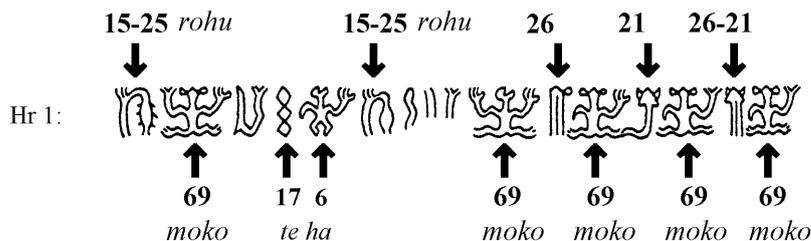

Figure 16.

Hr 1: **15-25 69 5-15 17 6 15-25 52 5-15 69 26 69 21 69 26-21 69**

*Rohu MOKO atua roa:* **te ha**! *Rohu, hiti: atua roa MOKO,* **ma** (or **mo-**) *MOKO ᵒko, MOKO* **ma-**(or **mo-**)*ᵒko, MOKO!*

Create (the words) 'LIZARD, the great god' **four (times)**! Create, lift (the words from this line to another): 'The great god LIZARD, **ma-** (or **mo-**) LIZARD ᵒ*ko*, LIZARD **ma-** (or **mo-**) ᵒ*ko*, LIZARD!'

Pupils wrote four lizard signs **69** *moko* together with the quasi-syllabic signs **26** *mo*, *ma*, *maa* and **21** *oko*, *ko* on lessons in the royal *rongorongo* school. The number of the lizard glyphs was brought out as the combination of glyphs **17 6** *te ha* (four).

In the light of this decipherment let us examine the text of the *manu* [bird] song from Easter Island:

*Kia teko te ha kia manau naunau a ure are rua te ha ka kai atu ka mau naunau no a ure are rua te ha* (Routledge 1914-1915).

I have reconstructed this text as follows:

*Kia Teko* **te ha**, *kia Ma. Naunaunau a ure are rua* **te ha**. *Ka kai atu, ka mau naunau no a ure are rua* **te ha**. To (the deity) *Teko* in the fourth (month), to (the deity) *Ma*. Using a piece of sandalwood, (hold) the flower (= the egg) (received) from a son (who took it) from the nest in the fourth (month). Hold (the egg), using (that) piece of sandalwood, hold firmly the flower (= the egg) (received) from the son (who took it) from the nest in the fourth (month).

Earlier I tried to decode this text, but I connected the term *naunau* with the name of the ceremonial platform Ahu Naunau at Anakena. But now I understand that the term *naunau* is the designation of a piece of sandalwood that was applied in the ceremony of the election of the new bird-man.

Métraux (1940: 336) says about the elected bird-man (employer) who received the sacred egg:

At Orongo the *hopu* handed the egg to his employer who had in the meantime shaved his hair, his brows, and his eyelashes. A priest (Routledge says a *rongorongo*, which is very unlikely) tied a strand of red tape and a piece of sandalwood (*ngaungau*) around the bird-man's arm which was tapu [taboo] from having received the egg. The priest recited a special incantation during the ceremony.

Rapanui *naunau* and *gnaungau* 'sandalwood' co-existed because of the alternation of the sounds *n/ng*. The form *naunaunau* is the incomplete doubling of the form *naunau*.

The magical significance of a piece of sandalwood is clear. The wordplay was possible: cf. Rapanui *ngau* 'to eat; to bite' and *kai* 'to eat; food.' So, the sacred egg was the symbol of the abundance of all kinds of the food.

A strand of red tape was the symbol of *Tiki-Makemake* because the red colour was the notation of that deity (see the details in Rjabchikov 2000).

The god *Teko* known as *Teko*-of-the-long-feet in a myth about *Tangaroa* was indeed the god *Rongo* (Ibid., pp. 310-311; Fedorova 1978b: 23), cf. also PPN *\*teki* 'to rise; to arrive.' The ceremonial village of Orongo (= *O Rongo*) was named after the god *Rongo* (Best 1924: 139). Some additional information about the early form of the god *Rongo* (*\*Logo*) has been obtained on the basis of the studies of Proto-Polynesian plots (Rjabchikov 2014: 164). *\*Teki*-of-the-long-feet (the swift *\*Logo*) was a mythological messenger of the ancient voyagers.

The last decoded text was namely that special incantation. The personage *Teko* was the god *Rongo*, and the personage *Ma* was the god *Makemake* (< *\*Ma-ke*), otherwise the god *Tiki* (*Tane*). The servant (young man) *hopu* (diver literally) was designated as *ure* (son). The specific term *are* (flower) denotes the egg here (cf. Rapanui *hua* 'fruit; flower'). Old Rapanui *hua* means 'egg' in some cases (Rjabchikov 2012b: 16-17). The phrase *ure are* correlates with the phrase *tama ere* (young man of an egg?) mentioned above in the chant "*Ka moe nga pua*." The Old Rapanui expression *kai atu/ata* (to receive gift; to hold in a hand) corresponds to the Rapanui expression *rima atakai* 'open-handed; gift,' cf. Maori *whakaatu* 'to show' and *kai* 'to reach; to arrive at.' The term *rua* (hole) means 'nesting place' (Barthel 1978: 152). The number four (*te ha*) denotes the month *Hora-nui* (September chiefly), namely the fourth month of the local calendar (cf. also Rjabchikov 2016a: 12-13, 15).

## Conclusions

The ancient priest-astronomers constantly watched many heavenly bodies. The record about Halley's Comet of 1682 A.D. has been decoded completely. Good agreement between it and the results of European astronomers is seen. The records about Halley's Comet of 1835 A.D. as well as about the sun, the moon, Mars and Saturn have been deciphered as well. The obtained information is the basis in order to understand some aspects of the bird-man cult.

## References

Alexander, J.D., 1981. Case-marking and Passivity in Easter Island Polynesian. *Oceanic Linguistics*, 20(2), pp. 131-149.
Barthel, T.S., 1957. Die Hauptgottheit der Osterinsulaner. *Jahrbuch des Museum für Völkerkunde zu Leipzig* 15, pp. 60-82.
Barthel, T.S., 1962. Rezitationen von der Osterinsel. *Anthropos*, 55(5/6), pp. 841-859.
Barthel, T.S., 1978. *The Eighth Land. The Polynesian Discovery and Settlement of Easter Island*. Honolulu: University of Hawaii Press.
Belyaev, N.A. and K.I. Churyumov, 1985. *Kometa Galleya i ee nablyudenie*. Moscow: Nauka.
Best, E., 1922. *The Astronomical Knowledge of the Maori*. Dominion Museum Monograph, vol. 3. Wellington: W.A.G. Skinner, Government Printer.
Best, E., 1924. *The Maori*. Vol. 1. *Memoirs of the Polynesian Society*. Vol. 5. Wellington: The Polynesian Society.
Best, E., 1925. The Burning of Te Arawa. *Journal of the Polynesian Society*, 34(136), pp. 292-320.
Blixen, O., 1973. Tradiciones pascuenses, II. Ure o Hei y los tres espíritus vengadores. Tuapoi. La vieja del brazo largo. La niña de la roto. *Moana, Estudios de Antropología Oceánica,* 1(6), pp. 1-11.
Campbell, R., 1999. *Mito y realidad de Rapanui: La cultura de la Isla de Pascua*. Santiago de Chile: Editorial Andrés Bello.
Chapin, P.G., 1978. Easter Island: A Characteristic VSO Language. In: W.P. Lehmann (ed.) *Syntactic Typology*. Sussex: Harvester Press, pp. 139-168.
Englert, S., 2002. *Legends of Easter Island*. Hangaroa: Rapanui Press/Museum Store.
Fedorova, I.K., 1978a. Nekotorye cherty istoricheskogo razvitiya rapanuyskogo yazyka. In: A.S. Petrikovskaya (ed.) *O yazykakh, fol'klore i literature Okeanii*. Moscow: Nauka, pp. 39-81.
Fedorova, I.K., 1978b. *Mify, predaniya i legendy ostrova Paskhi*. Moscow: Nauka.
Felbermayer, F., 1971. *Sagen und Überlieferungen der Osterinsel*. Nürnberg: Hans Carl.
Henry, T., 1928. Ancient Tahiti. *Bishop Museum Bulletin 48*. Honolulu: Bernice P. Bishop Museum.
Heyerdahl, T. and E.N. Ferdon, Jr. (eds.), 1965. *Reports of the Norwegian Archaeological Expedition to Easter Island and East Pacific*. Vol. 2. Miscellaneous Papers. Monographs of the School of American Research and the Kon-Tiki Museum, No 24, Part 2. Chicago – New York – San Francisco: Rand McNally.
Kronk, G.W., 1999. *Cometography: A Catalog of Comets*. Vol. 1: *Ancient – 1799*. Cambridge: Cambridge University Press.
Langdon, R. and D. Tryon. 1983. *The Language of Easter Island: Its Development and Eastern Polynesian Relationships*. Laie: The Institute for Polynesian Studies.
Lee, G., 1992. *The Rock Art of Easter Island. Symbols of Power, Prayers to the Gods*. Los Angeles: The Institute of Archaeology Publications (UCLA).
Liller, W., 1991. Hetu'u Rapanui: The Archaeoastronomy of Easter Island. In: P.M. Lugger (ed.) *Asteroids to Quasars: A Symposium Honouring William Liller*. Cambridge: Cambridge University Press, pp. 267-286.
McCoy, P.C., 1978. The Place of Near-shore Islets in Easter Island Prehistory. *Journal of the Polynesian Society*, 87(3), pp. 193-214.
Métraux, A., 1937. The Kings of Easter Island. *Journal of the Polynesian Society*, 46(2), pp. 41-62.
Métraux, A., 1940. Ethnology of Easter Island. *Bishop Museum Bulletin 160*. Honolulu: Bernice P. Bishop Museum.

Métraux, A., 1957. *Easter Island: A Stone-Age Civilization of the Pacific*. London: Andre Deutsch.

Mulloy, W., 1961. The Ceremonial Center of Vinapu. In: T. Heyerdahl and E.N. Ferdon, Jr. (eds.) *Reports of the Norwegian Archaeological Expedition to Easter Island and East Pacific*. Vol. 1. Archaeology of Easter Island. Monographs of the School of American Research and the Museum of New Mexico, No 24, Part 1. Chicago – New York – San Francisco: Rand McNally, pp. 93-180.

Mulloy, W., 1973. Preliminary Report of the Restoration of Ahu Huri a Urenga and Two Unnamed Ahu of Hanga Kio'e, Easter Island. *Bulletin 3*, Easter Island Committee. New York: International Fund for Monuments.

Mulloy, W., 1975. A Solstice Oriented *Ahu* on Easter Island. *Archaeology and Physical Anthropology in Oceania*, 10, pp. 1-39.

Popova, T., 2012. On One Samoan-Rapanui Lexical Parallel. *Polynesian Research*, 3(3), p. 8.

Randall, J.E. and A. Cea Egaña, 1984. Native Names of Easter Island Fishes, with Comments on the Origin of the Rapanui People. *Occasional Papers of Bernice P. Bishop Museum*, 25(12), pp. 1-16.

Rjabchikov, S.V., 1993. Rapanuyskie texty (k probleme rasshifrovki). *Etnograficheskoe obozrenie*, 4, pp. 124-141.

Rjabchikov, S.V., 1997. Easter Island Writing: Speculation and Sense. *Journal of the Polynesian Society*, 106(2), pp. 203-205.

Rjabchikov, S.V., 1998. Rapanui Placenames: Keys to the Mysteries. *NAMES: A Journal of Onomastics*, 46(4), pp. 277-281.

Rjabchikov, S.V., 1999. Astronomy and Rongorongo. *Rapa Nui Journal*, 13(1), pp. 18-19.

Rjabchikov, S.V., 2000. La trompette du dieu Hiro. *Journal de la Société des Océanistes*, 110(1), pp.115-116.

Rjabchikov, S.V., 2009. *Arkheologichesky pamyatnik Akahanga – Urauranga te Mahina na ostrove Paskhi*. Krasnodar: The Sergei Rjabchikov Foundation – Research Centre for Studies of Ancient Civilisations and Cultures.

Rjabchikov, S.V., 2012a. The rongorongo Schools on Easter Island. *Anthropos*, 107(2), pp. 564-570.

Rjabchikov, S.V., 2012b. The Teachers and Pupils on Easter Island. *Polynesian Research*, 3(3), pp. 9-25.

Rjabchikov, S.V., 2013a. *The Astronomical and Ethnological Components of the Cult of Bird-Man on Easter Island*. arXiv:1309.6056 [physics.hist-ph].

Rjabchikov, S.V., 2013b. *Some Notes on the Rapanui Archaeoastronomy*. arXiv:1311.0144 [physics.hist-ph].

Rjabchikov, S.V., 2014. The God Tinirau in the Polynesian Art. *Anthropos*, 109(1), pp. 161-176.

Rjabchikov, S.V., 2016a. *The Ancient Astronomy of Easter Island: Aldebaran and the Pleiades*. arXiv:1610.08966 [physics.hist-ph].

Rjabchikov, S.V., 2016b. *Tama-nui-te-ra*, *Tangaroa*, *Tane*, *Whiro*: Remarks on the Maori Pantheon. *Polynesia Newsletter*, 6, pp. 2-4.

Routledge, K., 1914-1915. *Katherine Routledge Papers*. Royal Geographical Society, London, Archives. Copies Held in Libraries and Research Centres.

Routledge, K., 1998. *The Mystery of Easter Island*. Kempton: Adventures Unlimited Press.

Thomson, W.J., 1891. Te Pito te Henua, or Easter Island. Report of the United States National Museum for the Year Ending June 30, 1889. *Annual Reports of the Smithsonian Institution for 1889*. Washington: Smithsonian Institution, pp. 447-552.